\documentclass[twocolumn,english,twocolum,superscriptaddress,bibnotes,amsmath,amssymb,floatfix,longbibliography]{revtex4-1}
\usepackage[colorlinks=true,citecolor=blue,linkcolor=black]{hyperref}
\usepackage[utf8]{inputenc}
\usepackage{lipsum}
\usepackage[english]{babel}
\usepackage{amsmath,amsfonts,amssymb}
\usepackage{mathtools}
\usepackage[T1]{fontenc}
\usepackage{url}
\usepackage{changes}
\usepackage{amsmath}
\usepackage{amsfonts}
\usepackage{amssymb}
\usepackage[version=3]{mhchem}
\usepackage{changes}
\definechangesauthor[color=red]{TJK}
\usepackage{epstopdf}
\usepackage{graphicx}
\graphicspath{{./Figures/}}

\begin{document}
\title{Gain-switched semiconductor laser driven soliton microcombs}

\author{Wenle Weng}
\email[]{wenle.weng@epfl.ch}
\affiliation{Institute of Physics, Swiss Federal Institute of Technology Lausanne (EPFL), CH-1015, Switzerland}

\author{Aleksandra Kaszubowska-Anandarajah}
\affiliation{CONNECT Research Centre, Dunlop Oriel House, Trinity College Dublin, D 2, Ireland}

\author{Jijun He}
\affiliation{Institute of Physics, Swiss Federal Institute of Technology Lausanne (EPFL), CH-1015, Switzerland}

\author{Prajwal D. Lakshmijayasimha}
\affiliation{Photonics Systems and Sensing Lab., School of Electronic Engineering, Dublin City University, Glasnevin, D 9, Ireland}

\author{Erwan Lucas}
\affiliation{Institute of Physics, Swiss Federal Institute of Technology Lausanne (EPFL), CH-1015, Switzerland}
\affiliation{Present address: Time and Frequency Division, NIST, Boulder, CO 80305, USA.}
\affiliation{Present address: Department of Physics, University of Colorado, Boulder, CO 80309, USA.}

\author{Junqiu Liu}
\affiliation{Institute of Physics, Swiss Federal Institute of Technology Lausanne (EPFL), CH-1015, Switzerland}

\author{Prince M. Anandarajah}
\email[]{prince.anandarajah@dcu.ie}
\affiliation{Photonics Systems and Sensing Lab., School of Electronic Engineering, Dublin City University, Glasnevin, D 9, Ireland}

\author{Tobias J. Kippenberg}
\email[]{tobias.kippenberg@epfl.ch}
\affiliation{Institute of Physics, Swiss Federal Institute of Technology Lausanne (EPFL), CH-1015, Switzerland}

\maketitle 
\noindent\textbf{\noindent
Dissipative Kerr soliton generation using self-injection-locked III-V lasers has enabled fully integrated hybrid microcombs that operate in turnkey mode and can access microwave repetition rates. Yet, continuous-wave-driven soliton microcombs exhibit low energy conversion efficiency and high optical power threshold, especially when the repetition frequencies are within the microwave range that is convenient for direct detection with off-the-shelf electronics. Here, by actively switching the bias current of injection-locked III-V semiconductor lasers with switching frequencies in the X-band and K-band microwave ranges, we pulse-pump both crystalline and integrated microresonators with picosecond laser pulses, generating soliton microcombs with stable repetition rates and lowering the required average pumping power by one order of magnitude to a record-setting level of a few milliwatts. In addition, we unveil the critical role of the phase profile of the pumping pulses, and implement phase engineering on the pulsed pumping scheme, which allows for the robust generation and the stable trapping of solitons on intracavity pulse pedestals. Our work leverages the advantages of the gain switching and the pulse pumping techniques, and establishes the merits of combining distinct compact comb platforms that enhance the potential of energy-efficient chipscale microcombs.
}

\section*{Introduction}
Developed nearly four decades ago, the gain switching technique is an elegant approach to generate optical pulse trains with semiconductor lasers \cite{tarucha1981response,van1982generation,paulus1988generation}. By applying a microwave signal to the bias current of a semiconductor laser, coherent pulses with a repetition rate up to a few tens of gigahertz can be emitted from the gain medium, constituting optical frequency combs \cite{zhou201140nm,anandarajah2011generation}. The simplicity, flexibility and low cost associated with this technique not only facilitate its use in classical optical telecommunications \cite{herbert2009discrete,imran2017survey} and secure quantum communication networks \cite{jofre2011true, yuan2014interference, yuan2016directly} but also may lead to valuable applications in microwave photonics \cite{marpaung2019integrated,tang2020photonics,cortes2020towards}. However, because of the intrinsic laser dynamics, the temporal durations of gain-switched laser (GSL) pulses are typically limited to a few tens of picoseconds, corresponding to narrow spectral spans of a few nanometers. Although spectral broadening with nonlinear fibre can be adopted to increase the GSL comb bandwidth \cite{anandarajah2011generation}, the need for multi-stage optical amplification and lengthy fibres precludes the use of GSLs in integrated broadband frequency comb generation.


\begin{figure*} [t]
\centerline{\includegraphics[width=1.67\columnwidth]{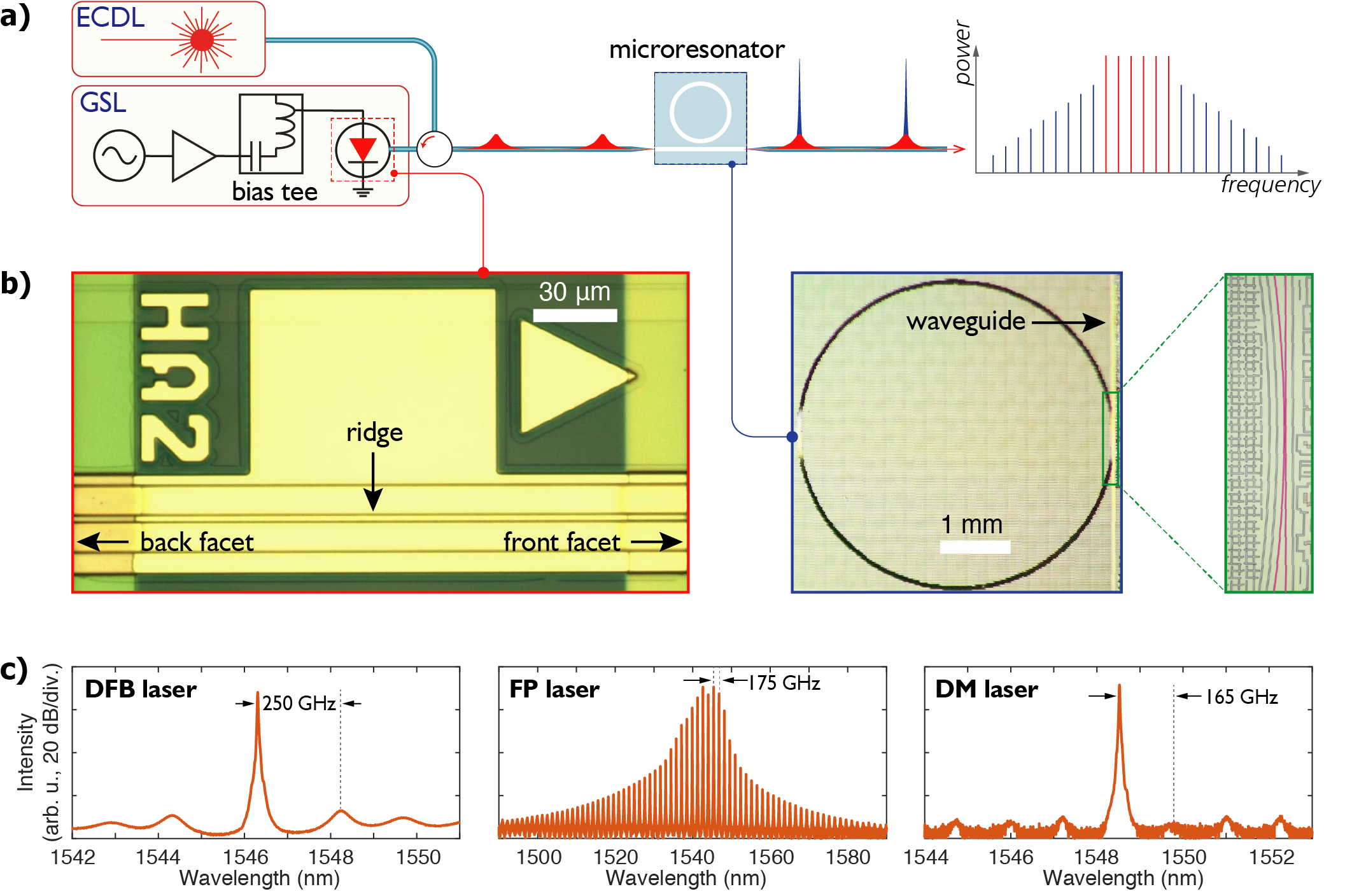}}
\caption{\textbf{Gain-switched semiconductor lasers for pulsed pumping of microcombs.} (a) Pumping scheme for soliton generation. The bias current of an injection-locked semiconductor laser is modulated to produce a pulse train to generate DKSs in a microresonator. A microwave synthesiser (Rohde\&Schwarz SMB100A) is providing the gain-switching frequency. (b) The photographs of (left) a Fabry-Perot laser chip, (middle) a Si$_3$N$_4$ microresonator, and (right) the enlarged coupling area (the microring and the waveguide are in pink for visibility). (c) The CW lasing spectra (i.\,e., without optical injection, gain switching deactivated) of the three semiconductor lasers used in this study, including (left) a distributed feedback laser, (middle) a Fabry-Perot laser; and (right) a discrete mode laser.
}
\label{fig1}
\end{figure*}

On a different path towards the development of fully integrated frequency combs, soliton microcombs \cite{kippenberg2018dissipative,gaeta2019photonic} have been revolutionising the frequency comb technology by producing femtosecond laser pulses with Kerr microresonators. Owing to the broad spectral width and the inherent low timing jitter of microresonator dissipative Kerr solitons (DKSs), besides the demonstration of coherent communications \cite{marin2017microresonator}, soliton microcombs have also been used in system-level applications in ultrafast ranging \cite{trocha2018ultrafast,suh2018soliton} and parallel Lidar \cite{riemensberger2020massively}. In the meantime, rapid progress in the development of soliton microcombs with ultralow power threshold or with powerful pump laser diode have been achieved \cite{stern2018battery,raja2019electrically,shen2019integrated,chang2020ultra}. In particular, advanced fabrication processes have enabled low-repetition-rate ($< 20$\,GHz) soliton excitation that is of crucial importance to integrated microwave photonics \cite{suh2018gigahertz,suh2019directly,liu2020photonic}. Yet, the generation of low-repetition-rate soliton microcombs with the traditional continuous-wave (CW) pumping scheme typically needs a pump power of at least a few tens of milliwatts \cite{suh2019directly,liu2020photonic}, which puts a stringent requirement on the tolerable loss of the system. To break away from the paradigm of CW pumping, the pulse-pumping technique has been demonstrated \cite{obrzud2017temporal} and applied in the calibration of astronomical spectrographs \cite{Obrzud:2019aa}. Instead of filling the entire resonator with intensive laser field, the pulse-pumping technique feeds in an intracavity field of soliton pedestal that occupies only a small portion of the cavity round trip, thus substantially lowering the power threshold and increasing the conversion efficiency. To date, however, this unconventional scheme is only realised by utilising cascaded high-speed modulators \cite{murata2000optical}, which requires bulky and expensive microwave components. As a result, the benefit of the reduced power threshold is to a large extent cancelled out by the high optical power loss induced by the modulators and the high radiofrequency (rf) driving power.

In this work we show that the GSL and the DKS can join together to complement each other and to bring about complex soliton dynamics. We apply the gain-switching technique on different semiconductor lasers including a distributed feedback (DFB) laser, a Fabry-Perot (FP) laser and a discrete mode (DM) laser, to demonstrate pulse-driven soliton microcombs in two microresonator platforms, namely a crystalline magnesium fluoride (MgF$_2$) whispering gallery mode resonator (WGMR) with a K-band microwave ($\sim$14\,GHz) repetition frequency ($f_{\rm rep}$) \cite{weng2019spectral} and a silicon nitride (Si$_3$N$_4$) integrated microresonator with an X-band microwave $f_{\rm rep}$ ($\sim$10\,GHz) \cite{liu2020photonic}. We show that the GSL pumping method allows for deterministic single soliton excitation and substantially lowers the optical power threshold. We also investigate the critical role of the phase of the GSL pulse in supporting soliton existence. We present two effective approaches to tailor the pulse profile in order to optimise robust microcomb generation, which not only paves the way to inexpensive integrated frequency combs but also reveals the fundamental soliton dynamics in a pulse driven scenario.

\begin{figure*} [t]
\centerline{\includegraphics[width=2.02\columnwidth]{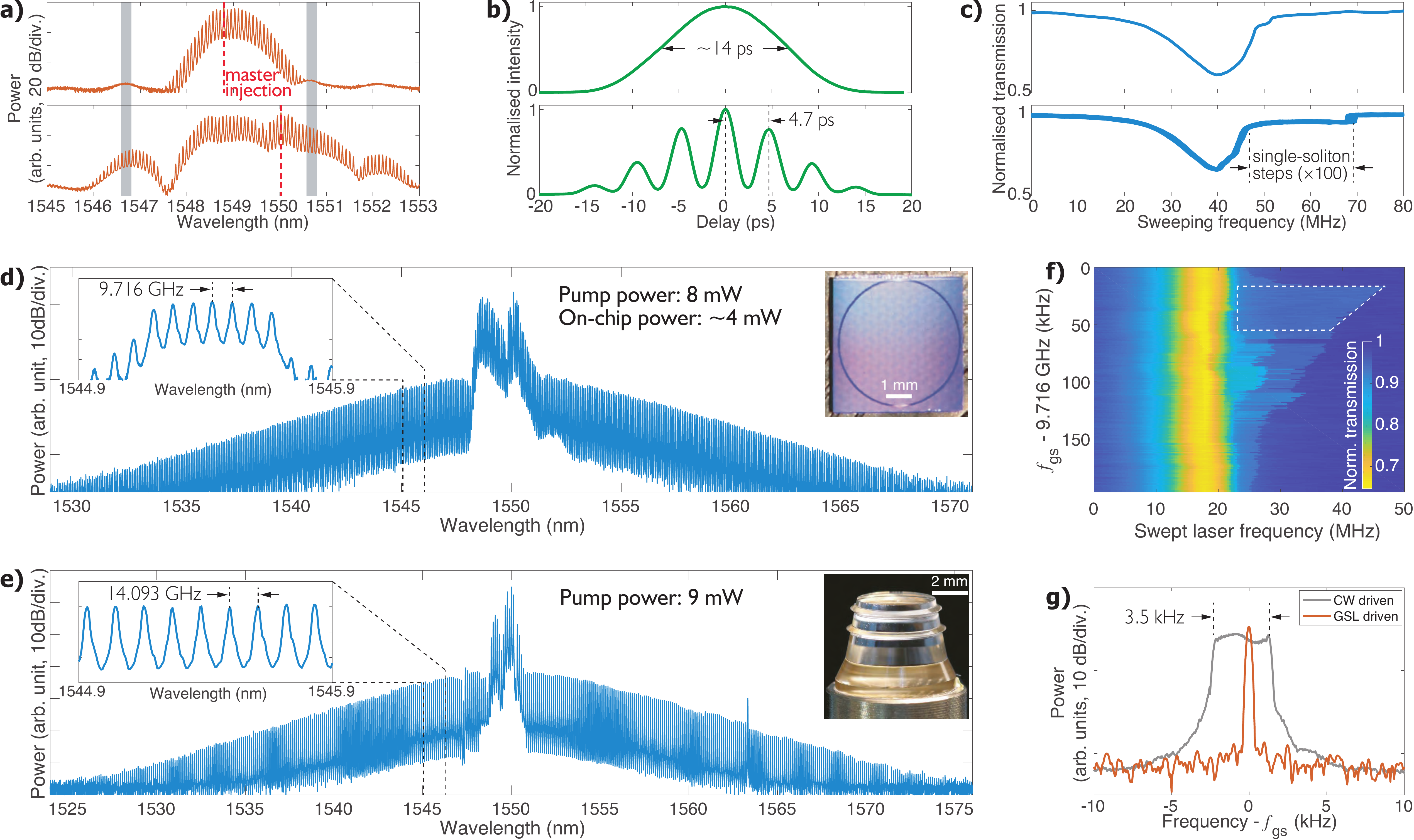}}
\caption{\textbf{Soliton microcomb generation with a gain-switched distributed feedback laser.} (a) The GSL spectra when the master laser is injected at the central lasing position (upper panel) and an off-central-mode frequency (lower panel). The grey stripes indicate the positions of the two adjacent longitudinal modes. The red dashed lines show the wavelengths of the master laser. (b) The intensity autocorrelation of the GSL pulses when the DFB laser is central-mode-injected (upper panel) and off-centre-injected (lower panel), respectively. (c) Laser down-sweeping transmission traces corresponding to the central-mode-injected operation (upper panel) and off-centre-injected operation (lower panel) respectively. In the lower panel 100 consecutive sweeping traces are overlaid. The offset-centre injection locking leads to significantly longer soliton steps. (d - e) The microcomb spectra of a Si$_3$N$_4$ microring resonator and a MgF$_2$ crystalline resonator. The left-hand insets show the enlarged portions of the spectra (bandpass-filtered for the Si$_3$N$_4$ microcomb). The right-hand insets are the photographs of the microresonators. (f) Contour plot of the laser-frequency-swept transmission as the gain switching frequency $f_{\rm gs}$ is varied. The trapezium outlined by dashed lines indicates the area for robust single-soliton generation. (g) Spectra of $f_{\rm rep}$ when the laser-resonance detuning is swept, showing that the microcomb repetition frequency is changing in CW-pumping but governed by $f_{\rm gs}$ in the GSL-driven situation. The measurement resolution bandwidth (RBW) is 100\,Hz.}
\label{fig2}
\end{figure*}

\section*{Results}
\noindent\textbf{Semiconductor lasers for microcomb generation.}
Figure\,\ref{fig1}\,(a) illustrates the principle of the GSL pumping technique. To pulse-pump a soliton microcomb, the bias current of a semiconductor laser is modulated at a frequency close to the microresonator free spectral range (FSR). Because the intrinsic linewidths of the semiconductor laser modules (without isolators) used in this study are of a few megahertz, we use a 1550-nm external cavity diode laser (ECDL) to injection-lock the semiconductor lasers, thus narrowing the linewidths of the slave lasers to $\sim20$\,kHz and ensuring the pulse-to-pulse coherence during gain switching. In the future this external injection locking setup may be replaced with self-injection locking using backscattering from the microresonator \cite{liang2015high,pavlov2018narrow}. Figure\,\ref{fig1}\,(b) shows the photographs of a gain-switched laser chip and a photonic microresonator, and Figure\,\ref{fig1}\,(c) presents the natural CW lasing spectra of the DFB laser, the FP laser and the DM laser, respectively. While the DFB laser shows single mode lasing behaviour, two first-order side modes approximately 250\,GHz apart from the central lasing mode are visible in the spectrum. In contrast, the FP laser exhibits multimode lasing with a mode separation of 175\,GHz. Here we particularly introduce the DM laser, which is essentially a FP laser with additional etched features to engineer the losses of different cavity modes to obtain single mode lasing. With a longitudinal mode spacing of 165\,GHz, the DM laser spectrum displays a high side-mode-suppression ratio (SMSR) of $\sim62$\,dB.

\noindent\textbf{Pulse driving with DFB GSL.}
We activate the gain switching of the DFB laser with a gain-switching frequency ($f_{\rm gs}$) provided by a microwave synthesiser. The upper panel in Fig.\,\ref{fig2}\,(a) shows a typical GSL spectrum (average power of 1\,mW) with $f_{\rm gs}$ of 10\,GHz when the ECDL is injected into the central mode of the DFB laser. We perform autocorrelation measurement of the amplified GSL pulses. The autocorrelation intensity displayed in the upper panel in Fig.\,\ref{fig2}\,(b) shows a width of $14$\,ps, indicating a pulse duration of $\sim10$\,ps (assuming a Gaussian profile). The master laser is swept with decreasing frequency over a resonance, while $f_{\rm gs}$ is finely adjusted around the expected $f_{\rm rep}$, in order to observe the step-like soliton signature in the transmission spectra. However, with an average GSL power of 120\,mW, which is even higher than the power level needed for CW driven soliton generation, we do not observe steps that are wide enough for stable soliton existence. The upper panel in Fig.\,\ref{fig2}\,(c) presents an example of the transmission spectra. With the optimised step width of only $\sim2$\,MHz during the laser sweep, we are not able to accurately stop the laser scan within the step range to enable soliton microcomb generation.

Surprisingly, when the ECDL is detuned from the DFB laser's central mode by more than 1\,nm, the GSL spectrum and the pulse profile change qualitatively. As can be seen from the lower panels in Fig.\,\ref{fig2}\,(a) and (b), the GSL comb bandwidth increases as the spectrum shows several lobes, and the autocorrelation intensity shows multiple peaks with a temporal separation of $\sim4.7$\,ps, suggesting the splitting of GSL pulses. With such an offset-injection-locked state, the GSL is able to generate soliton microcombs in both the Si$_3$N$_4$ and the MgF$_2$ microresonators with an average pump power below 10\,mW. Figure\,\ref{fig2}\,(d) and (e) show the generated Si$_3$N$_4$ soliton microcomb with $f_{\rm gs}=9.716$\,GHz and the MgF$_2$ soliton microcomb with $f_{\rm gs}=14.093$\,GHz. The pump powers at a-few-milliwatt level, to the best of our knowledge, are the lowest used in low-repetition-rate soliton microcombs. One should note that, due to the 3-dB lensed-fibre-to-waveguide loss, for the Si$_3$N$_4$ microcomb the on-chip optical power is below 5\,mW. Compared with prior works using CW pumping \cite{weng2019spectral,liu2020photonic}, the GSL pumping technique lowers the required pump power by an order of magnitude without compromising the spectral width of the soliton microcomb.

In the following we focus our effort on the Si$_3$N$_4$ microcomb. With a GSL power of 15\,mW, we vary $f_{\rm gs}$ as we repeat the laser sweep to produce the spectrogram in Fig.\,\ref{fig2}\,(f). In the area outlined with dashed lines single-soliton steps wider than 20\,MHz are observed. We set $f_{\rm gs}$ to be a constant value within this area, and perform the laser sweep for 100 consecutive times. The transmission traces are presented in the lower panel of Fig.\,\ref{fig2}\,(c), showing 100\% successful deterministic single soliton excitation rate. To further confirm the pulse pumping nature, we use an optical bandpass filter (bandwidth $\sim0.5$\,nm) to select the soliton microcomb at 1545.5\,nm for $f_{\rm rep}$ measurement. The filtered comb power is sent to a fast photodetector, whose output voltage is analysed by an electrical spectrum analyser (ESA). With the ECDL frequency scanned for 20\,MHz within the soliton step range, only a single repetition frequency is observed, showing that $f_{\rm rep}$ is disciplined by $f_{\rm gs}$. In contrast, with CW pumping, $f_{\rm rep}$ is changed by $\sim3.5$\,kHz due to Raman and higher-order dispersion effects that couple the soliton repetition rate to the pump-resonance detuning \cite{bao2017soliton,yi2016theory,karpov2016raman}. 

We deduce that the master laser injected at a position between the central mode and a side mode can induce the simultaneous lasing of both modes, and the interference between the two fields causes the multiple peaks in the GSL pulses. To test our hypothesis with numerical simulations, we develop a dual-lasing-mode model based on laser rate equations \cite{tarucha1981response,o2015numerical}, which are expressed as:
\begin{multline}
\label{eq1}
\frac{dN}{dt} = \frac{I_\mathrm{bias} + I_\mathrm{gs} \sin(2\pi f_\mathrm{gs} t)}{eV} - (\gamma_1 N + \gamma_2 N^2 + \gamma_3 N^3) \\
- aV(N-N_0) |A_\mathrm{GSL}|^2 + F_\mathrm{N}(t)
\end{multline}
\begin{multline}
\label{eq3}
\frac{dA_1}{dt} = \frac{1}{2} (1 - i \alpha_\mathrm{H}) \left(a V (N-N_0) - \frac{1}{\tau_\mathrm{p1}}\right) A_1 \\
+ \eta \sqrt{\kappa_\mathrm{c}} s_\mathrm{inj} + F_\mathrm{A}(t)
\end{multline}
\begin{multline}
\label{eq4}
\frac{dA_2}{dt} = \frac{1}{2} (1 - i \alpha_\mathrm{H}) \left(a V (N-N_0) - \frac{1}{\tau_\mathrm{p2}}\right)A_2 \\
+ i 2 \pi f_{\rm sm} A_2 + (1-\eta) \sqrt{\kappa_\mathrm{c}} s_\mathrm{inj} + F_\mathrm{A}'(t)
\end{multline}
where $N$ is the carrier density, $A_1$ and $A_2$ are the slowly varying envelopes of the fields of the central mode and the side mode respectively, $f_{\rm sm}$ is the frequency of the side mode relative to the central mode, $\alpha_\mathrm{H}$ is the linewidth enhancement factor, $a$ is the differential gain, $V$ is the active volume, $N_0$ is the carrier density at transparency, $\kappa_\mathrm{c}$ is the injection coupling rate, and $|s_\mathrm{inj}|^2$ is related to the injected photon flux. The carrier recombination rate is represented by $\gamma_1 N + \gamma_2 N^2 + \gamma_3 N^3$, showing the combined effects of the non-radiative recombination, the radiative recombination and the Auger recombination. The elementary charge is denoted by $e$. The photon density $|A_\mathrm{GSL}|^2$ in the laser active volume is related to $A_1$ and $A_2$ with $A_{\rm GSL} = A_1 + A_2$. For gain switching operation the gain-switching current $I_\mathrm{gs}$ is added to the bias current $I_\mathrm{bias}$. The spontaneous emission into the lasing field and the stochastic carrier recombination are denoted by the Langevin noise terms $F_\mathrm{A}$ and $F_\mathrm{N}$ respectively \cite{schunk1986noise,o2015numerical}. For simplicity, the ratio of the master laser field injected into the central mode ($\eta$) is set to be 0.5, and the two modes are assumed to have the same gain and linewidth enhancement factor, but different photon lifetimes. Moreover, this model neglects nonlinear interactions such as four-wave mixing. One should note that the DFB laser dynamics cannot be accurately described by this model as the modes in a DFB structure cannot be characterised with a photon lifetime like a FP mode. Nevertheless, we will show that this simplified model captures the key features of the dual-mode lasing phenomenon.

\begin{figure*} [htb]
\centerline{\includegraphics[width=1.9\columnwidth]{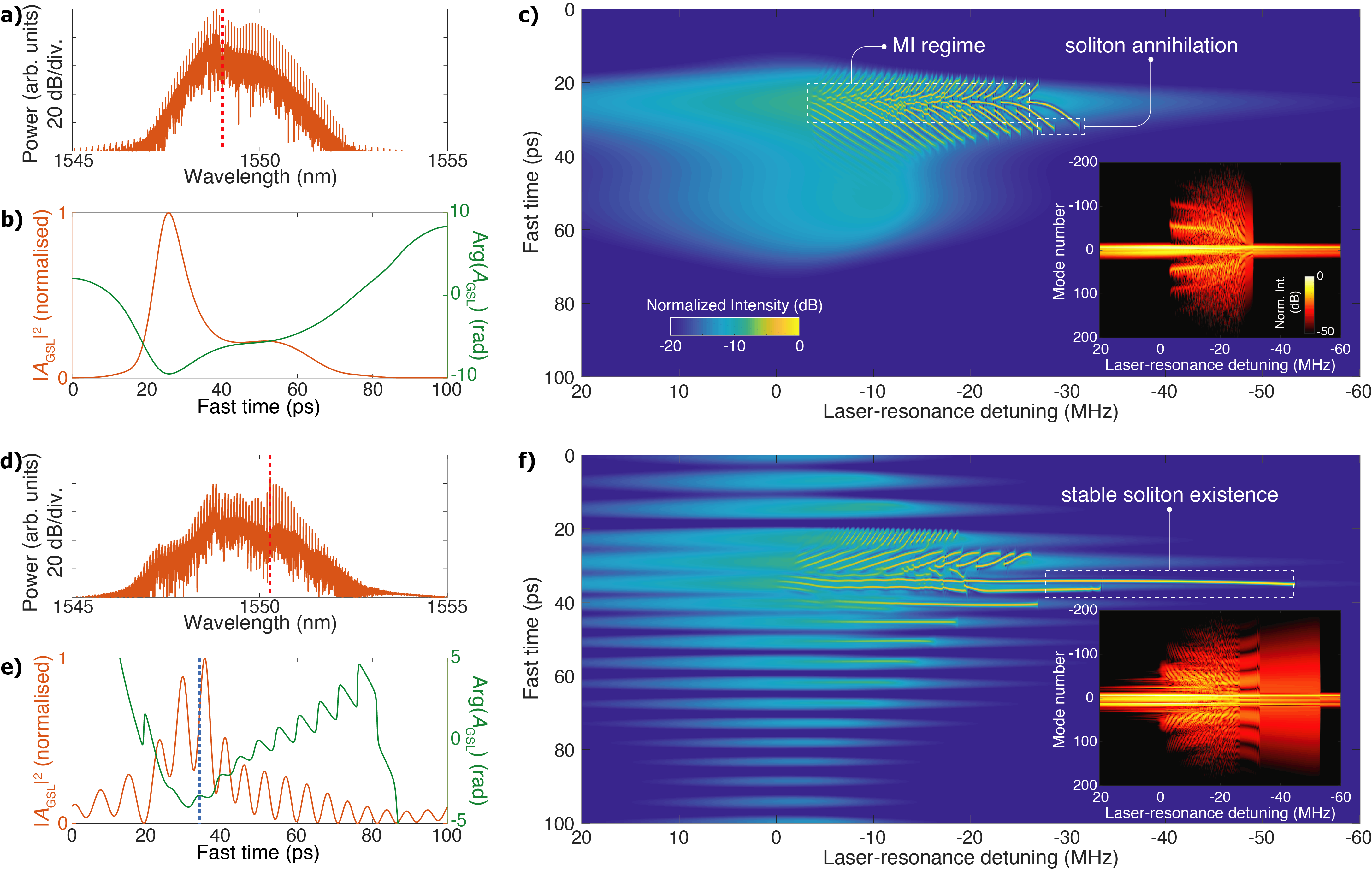}}
\caption{\textbf{Simulations of soliton generation with gain-switched laser driving.} (a) Comb spectrum and (b) pulse intensity and phase of the gain-switched laser when the master laser is injected into the centre of the lasing mode at 1549\,nm. The red dashed line indicates the position of the master laser injection. (c) Intracavity intensity evolution of a Si$_3$N$_4$ microring resonator driven by the GSL pulse in (b). MI: modulation instability. The inset shows the microcomb spectrum evolution. (d) Comb spectrum of the gain-switched laser that is offset-injection-locked. The master laser frequency is detuned from the centre of the intrinsic lasing mode by $-160$\,GHz. (e) Intensity and phase of the offset-injection-locked GSL pulse. The blue dashed line indicates the phase maximum that has the highest field amplitude. (f) Intracavity intensity evolution of the microring resonator that is pumped by the pulse in (e), showing a single-soliton step with a width of $\sim20$\,MHz yielded by a soliton trapped at the phase maximum that is indicated by the dashed line in (e).}
\label{fig3}
\end{figure*}

Figure \ref{fig3} (a) and (b) present the GSL spectrum and the intensity and phase (i.\,e., $\arg(A_{\rm GSL})$) of the pulse respectively when the master laser is injected at the centre of the natural lasing mode (see Methods for the details of the simulation), showing excellent agreement with the experimental results in terms of comb bandwidth and pulse duration. Crucially, the simulated pulse exhibits a phase minimum at the intensity maximum. Since DKSs are attracted to where the intracavity pumping field has the highest phase \cite{luo2015spontaneous,jang2015temporal,taheri2015soliton}, the GSL pulse profile is unfavourable to soliton generation because any soliton initiated by modulation instability (MI) \cite{godey2014stability} is drawn to the leading or the trailing edges of the pulse where the field amplitude is not adequate to support soliton existence. To confirm our analysis, we use the simulated GSL pulse as the pumping field profile to simulate microcomb generation using the Lugiato-Lefever equation (LLE) \cite{lugiato1987spatial,haelterman1992dissipative}. Neglecting higher-order dispersion but including Raman effect, the LLE is written as \cite{karpov2016raman}:
\begin{multline}
\label{eq5}
\frac{\partial{A}}{\partial t} + i \frac{D_{2}}{2} \frac{\partial^2{A}}{\partial \phi^2} - i g \left({|A|^2 A} - f_{\rm R} \phi_{\rm R} A \frac{\partial{|A|^2}}{\partial \phi} \right) =\\ 
\left( { - \frac{\kappa}{2} + i(\omega_0 - \omega_{\rm p}) } \right){A} + {\sqrt{\kappa _{\rm ex}} s_{\rm GSL}}
\end{multline}
where $A(\phi,t)$ is the intracavity field envelope, $\omega_0$ and $\omega_{\rm p}$ are the angular frequencies of the pumped resonance and the laser respectively. The definition of the other parameters and their values can be found in Methods and \cite{karpov2016raman}. We define the laser-resonance detuning as the frequency difference between the master laser and the corresponding resonance. As shown by the intracavity intensity evolution in Fig.\,\ref{fig3}\,(c), in the MI regime solitons are excited around the pulse peak. The formed solitons, however, are pulled to the pulse edges by the phase gradient, until they annihilate when the pumping field amplitude is below the minimum level required for soliton existence.

Next, we simulate the offset-injection-locked GSL by setting the master laser frequency to be 160\,GHz lower than the central mode frequency. With adjusted gain-switching current and injected power, our simulation results presented in Fig.\,\ref{fig3}\,(d) and (e) agree well with the experiments. This frequency-offset injection successfully excites the simultaneous lasing of two laser modes. While the gain-switched pulses from the two modes have distinct carrier frequencies, they overlap spatiotemporally as they are triggered by the same microwave signal. Remarkably, owing to the interference between the fields of the two constituent pulses, the dual-mode GSL pulse shows multiple intensity maxima and local phase maxima. These phase maxima, when their corresponding field amplitudes are strong enough, can trap solitons and facilitate robust soliton microcomb generation. Using the same procedure, we simulate the evolving intracavity field with the multipeak pulse profile, which is shown in Fig.\,\ref{fig3}\,(f). Indeed, after solitons are triggered by MI, they are stably confined by the phase maxima, despite that the Raman effect tends to shift solitons to locations with a larger fast time. As a result, a wide soliton step is obtained, enabling low-power-threshold DKS with an externally locked $f_{\rm rep}$.

\begin{figure*} [t]
\centerline{\includegraphics[width=1.46\columnwidth]{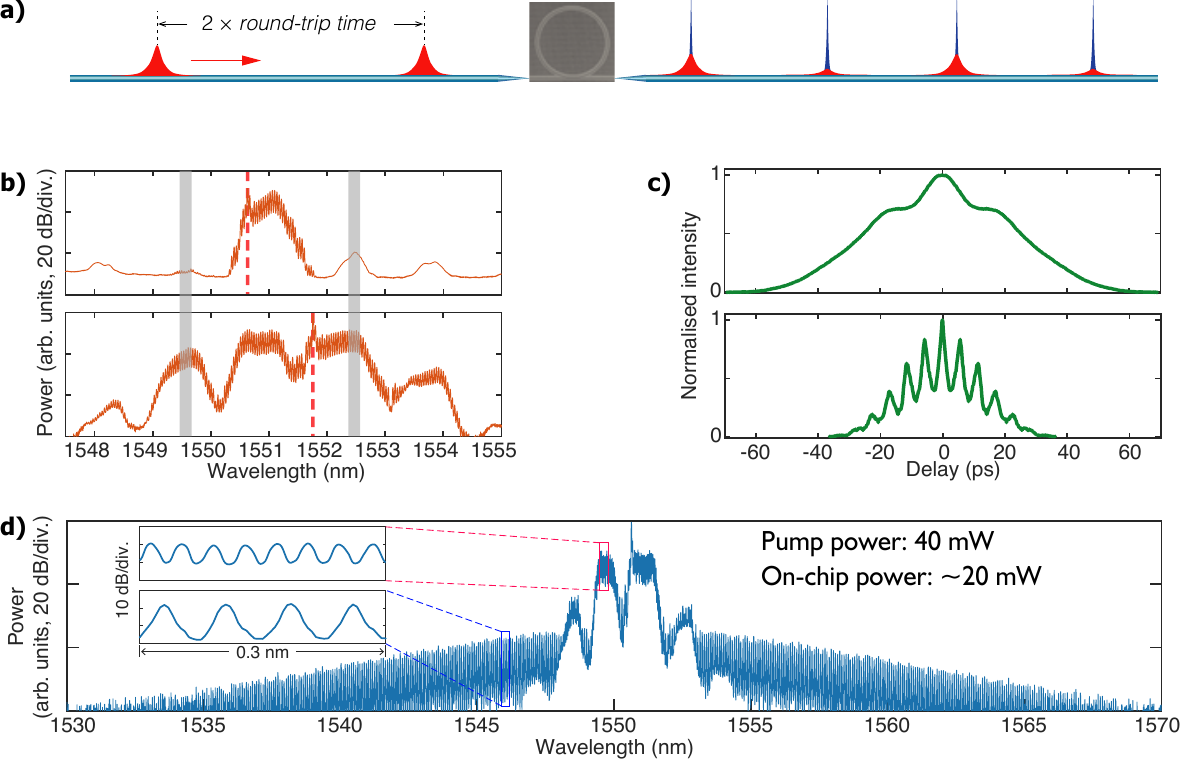}}
\caption{\textbf{Subharmonic pumping using a gain-switched Fabry-Perot laser.} (a) Schematic of the subharmonic pulse pumping method. (b) GSL spectra obtained with the master laser injecting around the centre of a lasing mode at 1550.7\,nm (upper panel) and between two lasing modes (lower panel), respectively. The grey stripes indicate the positions of the central mode's two neighbouring FP modes. The red dashed lines indicate the positions of the injection. (c) The measured autocorrelation traces of the GSL pulses in single-mode lasing (upper panel) and dual-mode lasing (lower panel) states, respectively. (d) The optical spectrum of the soliton microcomb driven by the offset-injection-locked FP GSL. The insets show the enlarged partial spectra of the GSL comb with $f_{\rm gs}$ of 4.86\,GHz and the microcomb with $f_{\rm rep}$ of 9.72\,GHz, respectively.}
\label{fig4}
\end{figure*}

\noindent\textbf{Subharmonic pumping with FP GSL.}
Since FP lasers are intrinsically multimoded, one would naturally expect the offset-injection-locking technique to be applicable with the FP GSL. Due to the limited gain-switching bandwidth of 7\,GHz of the FP GSL, we set $f_{\rm gs}$ to be half of $f_{\rm rep}$, which is around 4.858\,GHz. In this way we can pulse-pump the microresonator subharmonically (see Fig.\,\ref{fig4}\,(a)), with a halved pumping efficiency. The output power of the FP GSL is 4\,mW, and amplification is also used. The upper and the lower panels in Fig.\,\ref{fig4}\,(b) show the GSL spectra when the master laser is injected around the centre of a mode and offset-injected to excite multimode lasing, respectively. Again, offset-injection locking substantially expands the GSL comb bandwidth and creates multiple pulse peaks, shown by the comparison of the measured autocorrelation intensities in Fig.\,\ref{fig4}\,(c). While we still cannot observe any wide soliton steps with the single-mode-lasing GSL pulses with an average power up to 200\,mW, we can generate a single-soliton microcomb (whose spectrum is presented in in Fig.\,\ref{fig4}\,(d)) by using 40\,mW of the offset-injection-locked GSL. This power level is 4-fold larger than that for the DFB GSL because of the subharmonic pumping scheme and the CW component contained in the FP GSL. Nevertheless, the power threshold is lower than that of the CW pumping by nearly a factor of 3, showing the effectiveness and the flexibility of the GSL in exciting DKSs with repetition rates beyond the gain switching bandwidth.

\begin{figure*} [t]
\centerline{\includegraphics[width=1.7\columnwidth]{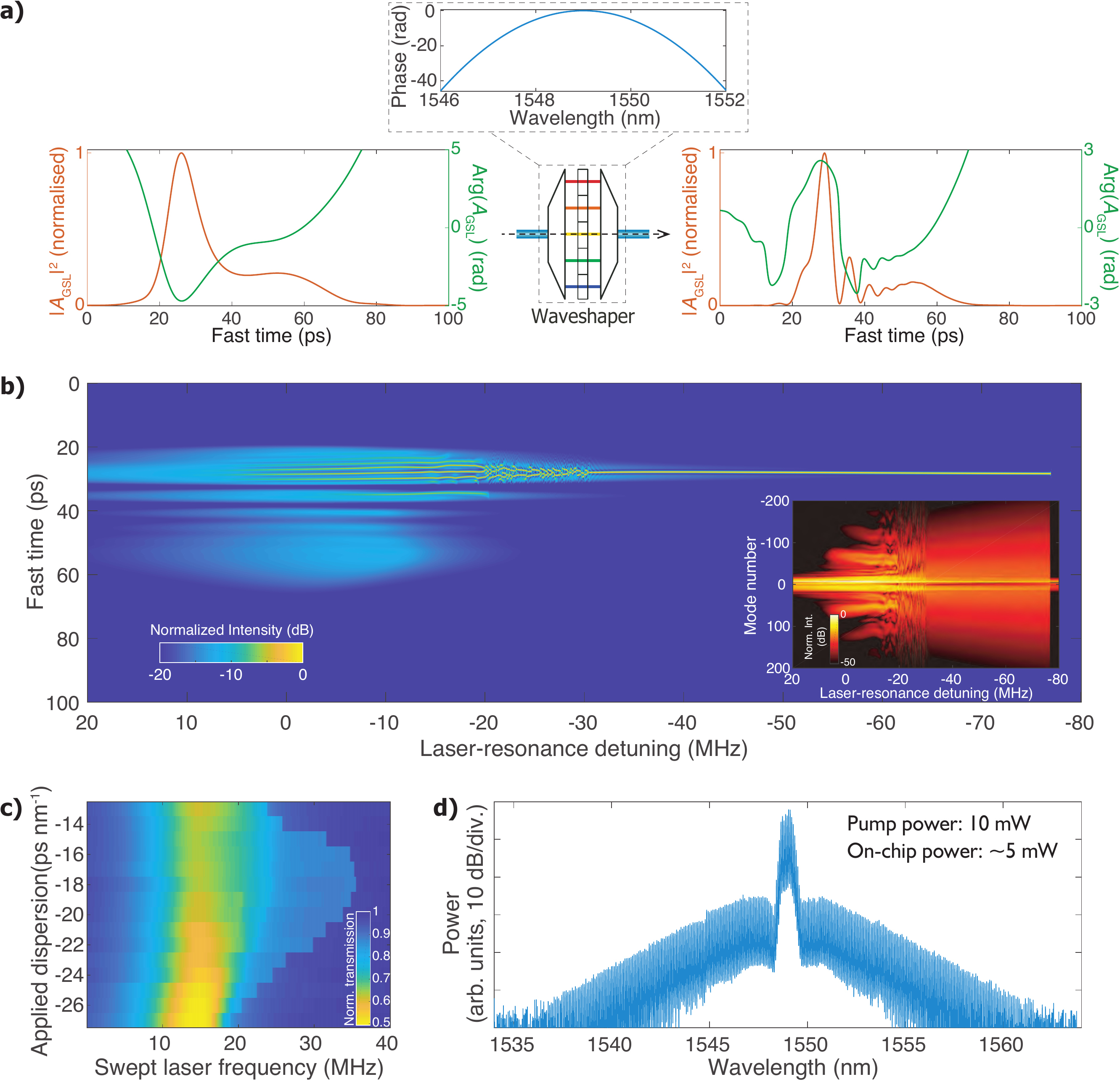}}
\caption{\textbf{Importance of phase engineering in pulse pumping with a gain-switched discrete mode laser.} (a) Auxiliary dispersion is applied to the simulated GSL pulse in order to achieve a phase maximum around the intensity peak. (b) Simulated soliton generation with externally chirped GSL pulses. The comb spectrum evolution is displayed in the inset. (c) Contour plot of the laser-frequency-swept transmission using the DM GSL with varied applied dispersion. (d) Single-soliton microcomb spectrum experimentally generated with 10\,mW of the DM GSL power using an external dispersion of $-18$\,ps\,nm$^{-1}$.}
\label{fig5}
\end{figure*}

\noindent\textbf{Pulse phase engineering with additional chirping.}
As recently reported in \cite{rosado2019numerical}, when the master laser is detuned from the lasing mode with a relatively large frequency offset, the DM GSL is not injection-locked anymore, and no coherent dual-mode lasing can be excited. This result, however, is not surprising given that the DM laser structure supports only a single lasing mode. To circumvent the difficulty in altering the phase of the DM GSL pulses, we use a waveshaper to apply external chirping. Figure \ref{fig5} (a) shows the computed pulse intensities and phases before and after the waveshaping with an applied dispersion of $-13$\,ps\,nm$^{-1}$. A local phase maximum is created by the waveshaping, and the pulse width simulated with single-mode laser rate equations is also narrowed. Inserting the phase-engineered pulse into the LLE model with an average pumping power of 5\,mW, we obtain a remarkably wide (>40\,MHz) single-soliton step (see Fig.\,\ref{fig5}\,(b)). Experimentally we apply chirping on the injection-locked DM GSL, and we tune the dispersion of the waveshaper to observe varied step widths displayed in Fig.\,\ref{fig5}\,(c). A dispersion of $-18$\,ps\,nm$^{-1}$ is found to yield the maximum step width, not only showing good agreement with the computation but also allowing us to generate a soliton microcomb with an average pump power of 10\,mW (see Fig.\,\ref{fig5}\,(d)). With a high biased current of 120\,mA, the averaged output power of the DM GSL is 13\,mW, which, in principle, could generate solitons in the Si$_3$N$_4$ microresonators without additional amplification. However, due to the significant ($>$4\,dB) loss induced by the waveshaper, amplification is still needed in this proof-of-principle experiment. Nevertheless, the demonstrated power threshold that is below the output power of the DM GSL shows promise for direct pulse-driven soliton generation if integrated dispersion compensation \cite{madsen1999integrated} is implemented.




\section*{Discussion}
A gain switching technique is implemented to generate low-repetition-rate soliton microcombs with semiconductor laser pulses. Phase alteration of the laser pulses, either using the dual-mode lasing state or applying auxiliary waveshaping, turns unfavourable intracavity laser field profiles into effective soliton traps that are necessary for the robust soliton generation and the flexible switching of microcomb states. The crucial role of the phase profile of the driving pulses could also have vital implications to soliton generation with advanced pulsed laser sources such as mode-locked quantum dot lasers \cite{rafailov2007mode}. Based on the fusion of two different compact frequency comb technologies, our work shows the prospect of using only chip-based components to generate energy-efficient wideband frequency combs, which could open the door for field-deployable applications of compact frequency combs in microwave photonics, sensing and metrology. Although the gain-switching scheme needs a microwave source, which potentially increases the complexity of the system, the flexible control and stabilisation of the soliton repetition rate is particularly beneficial to applications that require a fixed repetition rate and a rapidly tunable carrier frequency, such as frequency-modulated coherent Lidar \cite{riemensberger2020massively}.

Besides the significance to the development of practical microcomb sources, our work can be found useful for the study of complex laser dynamics \cite{yuan2016directly} and fundamental cavity soliton physics such as the spontaneous symmetry breaking and trapping of solitons by periodically modulated driving fields \cite{hendry2018spontaneous,taheri2017optical,xue2017soliton,cole2018kerr} and the nonlinear filtering effect in dissipative solitons \cite{brasch2019nonlinear}. 

\section*{Methods}

\noindent\textbf{Gain-switched semiconductor lasers.}
The DFB laser is manufactured by NTT Electronics. It is contained within a hermetically sealed high-speed package. The 7-pin package is incorporated with a high-speed rf input to enable direct modulation. The gain switching bandwidth of the laser is up to 20 GHz. The typical bias current is set between 40 and 60 mA.

The FP laser and the DM laser are commercially available from Eblana Photonics. They are packaged in high-speed butterfly packages. The FP laser has an effective gain-switching bandwidth of 7 GHz, and the applied bias current is around 60\,mA. The DM laser can be gain-switched with a frequency up to 14 GHz. It can be operated with a high bias current above 120 mA. Compared with the DFB laser, the DM laser shows a SMSR that is higher by 7\,dB, although its temperature-dependent wavelength tuning range is reduced to approximately 2\,nm, which is adequate for pumping microresonators with FSRs that are less than 250\,GHz.

To activate the gain-switching mode, a microwave tone derived from a microwave synthesiser (Rohde\&Schwarz SMB100A) is amplified and combined through a bias tee with the DC current. The power level of the microwave is typically between 18 and 26~dBm. To injection-lock the GSLs, an injected power around 1\,mW from the ECDL (Toptica Photonics) is sent into the GSLs via a fibre circulator. An erbium-doped fibre amplifier (Keopsys, maximum output power of 33~dBm) is utilised to amplify the GSL power for soliton pumping. We found that the pump laser linewidth usually needs to be much narrower than the microresonator mode bandwidth to reliably generate solitons, which is reasonable as only a narrower laser linewidth would allow the laser power to be efficiently accumulated in the resonator. This explains why compact semiconductor lasers usually need master laser injection locking or self-injection locking to assist in soliton microcomb generation. However, in order to fully understand the tolerable laser linewidth, a quantitative study on the laser frequency noise spectrum's impact on soliton stability is to be conducted in the future. 

\noindent\textbf{Microresonators.}
The MgF$_2$ WGMR is made from a z-cut crystalline rod with diamond shaping and surface polishing. A tapered fibre is used to couple light into the resonator. The loaded quality-factor ($Q$) of the mode resonance that generates combs is $\sim1\times10^9$, corresponding to a resonance bandwidth ($\frac{\kappa}{2 \pi}$) of 200\,kHz. The dispersion parameter ($D_2$) is approximately $2\pi\times2$ kHz.

The Si$_3$N$_4$ microresonator was fabricated with the Photonic Damascene Process \cite{pfeiffer2016photonic,liu2018ultralow}. The coupling section is designed to operate close to the critical coupling regime for 1550 nm. The loaded resonance bandwidth is around 20 MHz, corresponding to a loaded $Q$ of $1\times10^7$. The dispersion coefficient $D_2$ is measured to be $2\pi\times15$~kHz. Lensed fibres are used to couple light into and out of the nano-tapered waveguide \cite{liu2018double}.

\noindent\textbf{Gain-switched laser simulations.}
The numerical simulation is carried out with Runge-Kutta method. When the master laser injection is presented, the spontaneous emission is not necessary for simulating the GSL pulse profiles. However, we include this effect to obtain the GSL comb spectra with noise floors shown in Fig.\,\ref{fig3}. The noise terms are expressed as $F_\mathrm{A} = \sqrt{\beta \gamma_2 N^2 B_\mathrm{sim}} (e_\mathrm{I}(t) + i e_\mathrm{Q}(t))$ and $F_\mathrm{N} = \sqrt{2 (\gamma_1 N + \gamma_2 N^2 + \gamma_3 N^3) B_\mathrm{sim}} e_N(t)$. Here $\beta = 1\times10^{-4}$ is a factor related to the amount of spontaneous emission into the lasing mode, $B_\mathrm{sim} = 5\times10^{12}$\,Hz is the simulation bandwidth. The stochastic unit Gaussian random variable noise terms $e_\mathrm{N}$, $e_\mathrm{I}$ and $e_\mathrm{Q}$ have zero mean value and unity variance. For the optical field both the in-phase component $e_\mathrm{I}$ and the quadrature component $e_\mathrm{Q}$ are included.

For the GSL simulations, we use 4.5 for $\alpha_{\rm H}$, $4.5\times10^4$ s$^{-1}$ for $a$, $1\times10^{24}$ m$^{-3}$ for $N_0$, 2.5 ps for $\tau_{\rm P1}$, 1.4 ps for $\tau_{\rm P2}$ and $5\times10^{-17}$ m$^3$ for $V$. For the carrier recombination rate, $\gamma_1$ is $1\times10^9$ s$^{-1}$, $\gamma_2$ is $1\times10^{-16}$ s$^{-1}$m$^3$, and $\gamma_3$ is $1\times10^{-41}$ s$^{-1}$m$^6$. To simulate the central-mode-injected GSL dynamics, $I_{\rm bias}$ and $I_{\rm gs}$ are both set to be 75 mA, and the source term $\sqrt{\kappa_\mathrm{c}} s_\mathrm{inj}$ is $4\times10^{21}$ m$^{-3/2}$ s$^{-1}$. For the offset-injection-locking scenario, $I_{\rm bias}$ is 75 mA, $I_{\rm gs}$ is 55 mA, and $\sqrt{\kappa_\mathrm{c}} s_\mathrm{inj}$ is $3.28\times10^{22}$ m$^{-3/2}$ s$^{-1}$. We note that the set of parameter values we use are not for a particular GSL type in the experiments. Instead, they are chosen to be within the reasonable ranges of typical parameter values so the model is general and qualitative. For example, the linewidth enhancement factor (also known as the Henry factor) is set to be 4.5 because usually it is between 2 and 7, and for the DFB laser it was measured to be around 5 \cite{anandarajah2011generation}. To qualitatively reproduce the gain-switching behaviours, one can vary the values of multiple parameters and obtain similar results.

Because it is the relative phase of the pumping field that matters in the soliton pumping, all the simulated GSL phase profiles presented in this work are offset-adjusted to be centred around the 0 radian level.

\noindent\textbf{Pulse-driven soliton generation simulations.}
Some parameters in Eq.\,\ref{eq5} are defined as such: $D_{2}$ is the second-order dispersion coefficient, $\phi$ is the co-rotating angular coordinate that is related to the round-trip fast time coordinate $\tau$ by $\phi = \tau \times D_1$ (where $\frac{D_1}{2 \pi}$ is the FSR), ${\kappa_{\rm ex}}$ is the external coupling rate, ${\kappa}$ is the cavity decay rate that includes both the external coupling rate and the intrinsic loss rate caused by mechanisms such as material absorption and roughness-induced scattering (which is the dominant intrinsic loss mechanism in the microresonators in this work), and $s_{\rm GSL}$ is the input GSL field. In the LLE $g$ is the single photon induced Kerr frequency shift, $f_{\rm R}$ is the Raman fraction, and $\phi_{\rm R}$ is the term related to the Raman shock time \cite{karpov2016raman}.

To simulate the soliton generation with Eq.\,\ref{eq5}, the GSL pulse profiles are used as the pumping term after proper interpolation is performed to match the temporal resolution of the LLE simulation. The FSR of the resonator is set to be 10\,GHz, the second order dispersion coefficient $D_2$ is $2\pi\times15$\,kHz, and $g$ is 0.19\,Hz. the cavity decay rate is $\kappa = 2\pi\times20$\,MHz, and $\kappa_{\rm ex}$ is set to be equal to $\frac{\kappa}{2}$. To include the Raman effect, $f_{\rm R} = 0.008$ and $\phi_{\rm R} = 2\times10^{-4}$ rad are used. For all the simulations shown in Fig.\,\ref{fig3} and Fig.\,\ref{fig5}, the laser sweeping speed is $4\times10^{12}$\,Hz s$^{-1}$.

\noindent\textbf{Autocorrelation measurement.}
To measure the autocorrelation intensities of the GSL pulses, the GSL light is split interferometrically and recombined with an adjustable delay in a barium borate crystal to generate second harmonic signal in a non-collinear fashion. The intensity of the second harmonic signal is continuously measured as the delay is scanned. Due to the limited maximum delay of the setup, the measurement can only cover more than half of the full delay range for the FP GSL pulses whose repetition rate is relatively low. Thanks to the symmetrical nature of the autocorrelation traces, we horizontally flip the partial autocorrelation traces with respect to the positions of the intensity maxima to produce the full traces shown in Fig.\,\ref{fig4}\,(c).

\medskip
\noindent\textbf{Authors contributions}

\noindent W.\,W., A.\,K.-A. and P.\,M.\,A. conceived the project. W.\,W. performed the experiments with A.\,K.-A., J.\,H., and E.\,L.'s assistance. W.\,W. analysed the data, developed the theoretical model, and performed the numerical simulations. A.\,K.-A., P.\,D.\,L. and P.\,M.\,A. developed and tested the gain-switched lasers. J.\,L. designed and fabricated the silicon nitride resonator. J.\,H. and J.\,L. characterised the silicon nitride resonator. W.\,W. wrote the manuscript, with input from other authors. P.\,M.\,A. and T.\,J.\,K. supervised the project.
\medskip

\noindent\textbf{Competing interests}

\noindent The authors declare no competing interests.
\medskip

\noindent\textbf{Acknowledgments.}

\noindent W.W. thanks Romain Bouchand, Miles Anderson, Arslan Raja and Johann Riemensberger for technical assistance, and Cherise M. Qiu for assistance in illustration. The authors acknowledge the assistance from Rui Ning Wang on silicon nitride resonator fabrication. This publication was supported by Contract No.\,D18AC00032 (DRINQS) from the Defense Advanced Research Projects Agency (DARPA), Defense Sciences Office (DSO), funding from the Swiss National Science Foundation under grant agreement No.\,192293, Science Foundation Ireland (SFI) 15/CDA/3640 and the SFI/European Regional Development Fund (13/RC/2077). This work is also supported by the Air Force Office of Scientific Research under award number FA9550-19-1-0250.

%
\end{document}